\documentclass[journal]{IEEEtran}

\ifCLASSINFOpdf
\else
\fi
\usepackage[table]{xcolor}
\usepackage{amsmath,graphicx}
\usepackage{url}
\usepackage{amsfonts}
\usepackage{float}
\usepackage{graphicx}
\usepackage{multirow}
\usepackage{subcaption}
\usepackage{xcolor}
\usepackage[pagebackref=true]{hyperref} 
\renewcommand*{\backrefalt}[4]{%
     \ifcase #1 \footnotesize{(Not cited.)}%
     \or        \footnotesize{(Cited on page~#2)}%
     \else      \footnotesize{(Cited on pages~#2)}%
     \fi}\usepackage{cite}
\usepackage{algorithm}
\usepackage{algpseudocode}
\graphicspath{{Images/}}
\usepackage{tikz}
\def\checkmark{\tikz\fill[scale=0.4][color=black!60!green](0,.35) -- (.25,0) -- (1,.7) -- (.25,.12) -- cycle;}

\usepackage{booktabs}
\newcommand{\duf}{D$^\text{2}$UF}
\begin{document}

\title{D$^\text{2}$UF: Deep Coded Aperture Design and Unrolling Algorithm for Compressive Spectral Image Fusion}%

\author{Roman~Jacome,~\IEEEmembership{Student Member IEEE}
        Jorge~Bacca,~\IEEEmembership{Member IEEE}
        and~Henry~Arguello,~\IEEEmembership{Senior Member IEEE}
\thanks{R. Jacome is with the Department
of  Physics, Universidad Industrial de Santander, Colombia}
\thanks{J. Bacca and H. Arguello are with the Department of Computer Science, Universidad Industrial de Santander, Colombia.}
\thanks{This material is based upon work supported by the Air Force Office of
Scientific Research under award number FA9550-21-1-0326. }}


\maketitle
\IEEEpeerreviewmaketitle
\begin{abstract} 

Compressive spectral imaging (CSI) has attracted significant attention since it employs synthetic apertures to codify spatial and spectral information, sensing only 2D projections of the 3D spectral image. However, these optical architectures suffer from a trade-off between the spatial and spectral resolution of the reconstructed image due to technology limitations. To overcome this issue, compressive spectral image fusion (CSIF) employs the projected measurements of two CSI architectures with different resolutions to estimate a high-spatial high-spectral resolution.  This work presents the fusion of the compressive measurements of a low-spatial high-spectral resolution coded aperture snapshot spectral imager (CASSI) architecture and a high-spatial low-spectral resolution multispectral color filter array (MCFA) system. Unlike previous CSIF works, this paper proposes joint optimization of the sensing architectures and a reconstruction network in an end-to-end (E2E) manner. The trainable optical parameters are the coded aperture (CA) in the CASSI  and the colored coded aperture in the MCFA system, employing a sigmoid activation function and regularization function to encourage binary values on the trainable variables for an implementation purpose. Additionally, an unrolling-based network inspired by the alternating direction method of multipliers (ADMM) optimization is formulated to address the reconstruction step and the acquisition systems design jointly. Finally, a spatial-spectral inspired loss function is employed at the end of each unrolling layer to increase the convergence of the unrolling network. The proposed method outperforms previous CSIF methods, and experimental results validate the method with real measurements. 
 
\end{abstract}


\IEEEpeerreviewmaketitle

\section{Introduction}\label{sec:introduction}
Compressive spectral imaging (CSI) allows the estimation of the spatio-spectral information from coded and multiplexed projections \textit{i.e,} CSI does not directly sense the spectral information \cite{CSI}. CSI uses new optical systems that encode spatial or spatial-spectral information using synthetic apertures (SA). Then all the encoded spatial-spectral information is integrated into a monochromatic or RGB sensor. After obtaining the compressed measurements, a recovery algorithm is employed to estimate the spectral image. SA plays an essential role in CSI since it directly affects the structure of the measurements, similar to phase imaging or computer tomography.~\cite{vouras2022overview}. Consequently, the quality of the estimated spectral image primarily depends on the sensing procedure and the recovery method. 

In the sensing process, there is a trade-off between spatial and spectral resolution due to technological limitations~\cite{fusion_book,deepfusion}. For instance, a well-studied CSI architecture known as the multispectral color filter array (MCFA) system employs a SA called colored-coded aperture (CCA) to modulate the spectral and spatial information of the scene through the spectral responses of a set of filter spatially distributed~\cite{multispectral}. Narrowband color filters increase the spectral resolution, but compromise either the spatial resolution~\cite{mcfa_resolution1} or the acquisition rate, requiring a multi-shot sensing~\cite{mcfa_resolution2}.  Another well-known CSI architecture is the coded aperture snapshot spectral imaging (CASSI)~\cite{CASSI}. This acquisition system is based on a SA known as coded aperture  (CA), that modulates spatial information by blocking or unblocking certain scene regions. Then a prism decomposes the coded incident light into its spectral components, and finally, a monochromatic sensor integrates the coded and dispersed information. The use of dispersive elements allows for the acquisition of richer spectral information constraining the spatial resolution to FPA size~\cite{resolution}. 

Most spectral imaging applications, such as classification or anomaly detection~\cite{yo,hinojosa2018coded,karthik2021detection} require high spatial and spectral resolution. The fusion of two CSI with different resolutions has been introduced to overcome the resolution issue~\cite{fusion_book}. Then, a fusion-reconstruction algorithm is employed, which takes the features of both compressive measurements to obtain a high-spatial high-spectral resolution image, this methodology is called compressive spectral image fusion (CSIF). For instance, the authors of~\cite{CSIF1} employ the fusion of compressed data from two CASSI architectures with different resolutions {as well as the fusion} of two spatial-spectral encoded compressive spectral imagers (SSCSI)~\cite{CSIF1}, proposing an alternating direction of multipliers (ADMM) algorithm with sparsity and total variation priors. The work in \cite{CSIF2} proposes a non-local low-rank abundance prior to using two colored CASSI as sensing architecture also with ADMM algorithm. Also, the work in \cite{CSIF6} proposes an unrolled deep neural network inspired by a linearized ADMM algorithm for the fusion of two MCFA architectures, one with high-spectral but low-spatial-resolution ,and the other with low-spectral but high-spatial-resolution. The authors of~\cite{CSIF7} presented the first testbed implementation of a CSIF system that uses a digital micromirror device (DMD) to codify and split the incident light into two CSI arms; one senses high detailed spatial information, and the other senses high spectral resolution data. Also, {another method for fusion that has been proposed involves employing} a side sensor such as an RGB camera and a CSI system~\cite{CSIF5, CSIF4}. Additionally, the fusion of two CSI architectures has been used for a more accurate spectral image classification \cite{fusion_class,fusion_class2}. The aforementioned fusion systems use random codings and do not exploit the optimal design of {these optical elements}.

On the other hand, the quality of the spectral image depends on the recovery algorithm. Namely, there are model-based algorithms~\cite{lr} based on convex optimization, deep-learning methods~\cite{lambda_net,meng2020end,monroy2021deep} or model-inspired deep neural networks \cite{unr1,unr2,unr3}, the latter being the state-of-the-art in CSI. In addition to the networks, the loss function plays an important role in the network performance for CSI. Until now, the $\ell_1$ or the $\ell_2$ norm \cite{dnu} are used directly or in adversarial manner \cite{cheng2022recurrent} for training the reconstruction networks. However, these loss functions do not directly focus on highlighting the spectral fidelity, {which is the distinctive feature of spectral imaging}. Moreover, the hardware optical parameters, such as CA or CCA, can also be optimized following some design criteria to improve the condition of the sensing matrix and therefore improve the reconstruction quality. Several design criteria and methods have been proposed,  \cite{design1}, for instance, uses spatio-temporal correlation, or \cite{colored} use concentration of measure to satisfy the restricted isometry property (RIP), which defines the number of projections required for correct reconstruction. On the other hand, the End-to-End (E2E) optimization for CA introduced in \cite{bacca2021deep} leverages large datasets to jointly optimize  the CA  and the decoder operator, which performs a high-level task \textit{e.g.} classification, segmentation, or recovery. This coupled methodology models the sensing process as a layer of a deep neural network (DNN) where its trainable parameters are the CA, whose training is constrained to obtain implementable values such as grayscale \cite{gray}, or binary \cite{CSI}.
Particularly, the design of the coding elements in the MCFA and the CASSI architectures has attracted significant attention. The spatial distribution and spectral response of the CCA in the MCFA architecture have been widely studied \cite{multispectral} for efficient demosaicking. For the CASSI system, several design criteria have been used based on compressed sensing theory, such as RIP  \cite{CSI}. However, the joint design of the sensing parameters and the reconstruction algorithm for CSIF have not been exploited. 

Thus, an unrolled network for compressive spectral image fusion from CASSI and MCFA measurements and the joint design of CA and the CCA using an E2E approach is proposed. The learning of optical parameters (CA and CCA) constrains the E2E training since it needs to converge to implementable {values. This paper addresses this issue} by the inclusion of a regularizer function \cite{bacca2021deep} and a sigmoid activation function to promote implementable binary values. Furthermore,  since the unrolling network can be seen as an iterative process, we propose multiple losses for each unrolling iteration which accelerates the convergence and gives a practical guide on how many iterations are required for the fusion problem. Extensive simulations highlight improvements with respect previous CSIF works, and experimental results validate the proposed coded aperture design compared with blue-noise design \cite{design1} and random patterns for real data reconstruction. 

\section{Related Work}

\subsection{Coded aperture design}
Traditionally, CA designs are based on exploiting theoretical properties of compressive sensing such as RIP \cite{CSI, design1}, minimizing the zero singular values of the sensing matrices to promote uniform sampling of the signal \cite{design2}, optimizing the concentration of measure to reduce the number of projection required \cite{design3}. However, these methods are independent of the reconstruction algorithm. Recently, the E2E approach is a coupled methodology where the optical free-parameters are learned together with the weights of a DNN to perform a high-level task \textit{e.g.} reconstruction or classification, among others \cite{sitzmann2018end}. For instance, \cite{designdl} proposes the joint learning of the CA of the CASSI architecture with a reconstruction network. To constrain the CA entries to binary values, the authors employ a sign function that brings problems in the back-propagation process because its derivative has an infinite value at 0. In \cite{design_ca}, the CA aperture is learned to improve the quality of the classification in a single-pixel camera. Here the binary constraint of the CA is imposed as a differentiable regularization in the loss function. Similarly, \cite{bacca2021deep} proposes a family of regularizes to exploit other CA properties such as temporal correlation, transmittance, and the number of shots, among others, for reconstruction, classification, and semantic segmentation tasks. Neither traditional CA design nor E2E approach has been explored for CSIF.  

\subsection{Compressive spectral image reconstruction}

With advances in deep learning approaches for inverse problems \cite{inverse}, several deep learning-based reconstruction methods have been proposed for CSI. For instance, \cite{lambda_net} uses a two-stage DNN, a reconstruction stage that uses self-attention mechanisms along with a discriminator, and a refinement stage employing a U-net architecture. {Another example is  \cite{deeprecon1}, which proposes} a deep convolutional neural network (DCNN) to extract feature maps of the projection and a residual operation of the reconstruction. In these approaches, the reconstruction process lacks interpretability and flexibility, since the DNN works as a black-box model, { which causes problems in generalization}. To address this issue but also exploit the use of a large dataset to improve the model, the unrolling networks (UN) have been proposed \cite{unrolling_review}. In UN,  the layers of the DNN perform an iteration of an iterative algorithm, allowing a model-based interpretation of the network. In CSI, several UN {have achieved remarkable results, like that presented in}  \cite{dnu}, which proposes nonlocal-local data-driven prior constructed with convolutional layers in half quadratic splitting algorithm,  or the method presented in \cite{sogabe2020admm}, which unrolls the iterations of an ADMM algorithm in DNN. However, these methods use random CA, which suffers from low-conditionality of the sensing matrix.

\subsection{Compressive spectral image fusion methods}

Several computational algorithms for CSIF have been proposed. For instance, the works in \cite{CSIF5} and \cite{CSIF4} elaborate a non-iterative approach to fuse compressive spectral measurements with a RGB image. In \cite{CSIF1, CSIF2, CSIF8} the fusion and reconstruction process is performed with an ADMM algorithm using prior information such as sparsity, total variation, non-local low-rank abundance, or sparse representation of abundance maps. The authors in {\cite{CSIF8}} formulates a UN for a linearized ADMM algorithm for the fusion problem. Nevertheless, these methods do not exploit the design of the CA to improve the performance of the reconstruction algorithm. 
\subsection{Contributions of this work}
The main contributions of the framework are summarized in the following items: 
\begin{enumerate}
    \item An E2E optimization for CSIF: We propose an optimal design of the CA in CASSI and the CCA in the MCFA system based on E2E training of the optical parameters and the reconstruction network.
    \item CSIF unrolling network: An optimization-based DNN is proposed for CSIF where every stage of the network performs an ADMM iteration
    \item Multiple-loss strategy: Towards an improvement in the convergence of the reconstruction network along the stages, we propose the use of the multiple-loss at the end of each unrolling stage to avoid gradient vanishing in the first stages, which allows one to experimentally set the optimal number of stages in the unrolled network.
    \item Loss function: We employed a loss function promoting the visual enhancement of the reconstructed image and an improvement of the spectral signature reconstruction.  
\end{enumerate}

\begin{figure}[!t]
    \centering
    \includegraphics[width=0.45\textwidth]{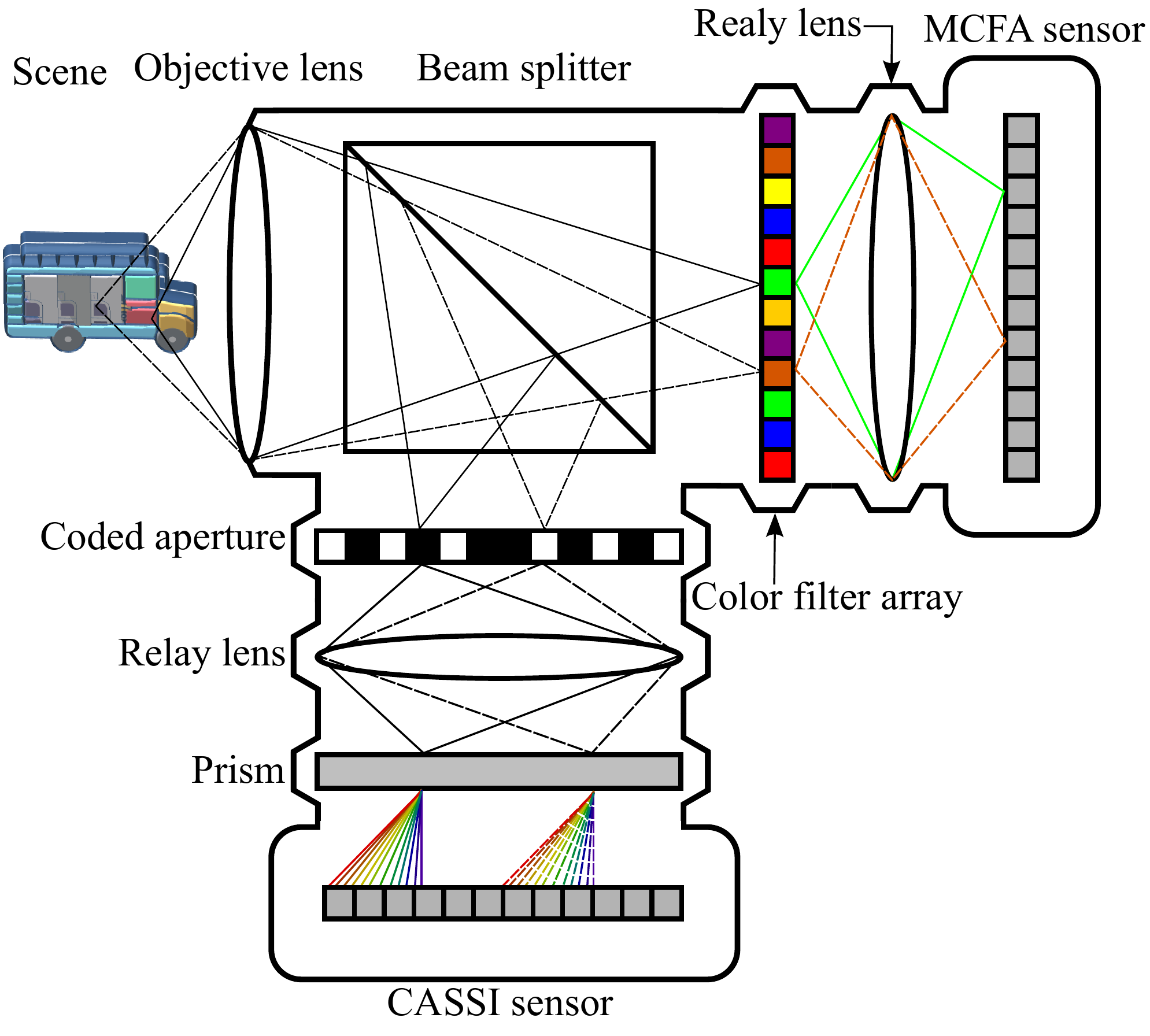}
    \caption{Proposed CSIF system based in the CASSI and MCFA architectures. The input light source is divided in two by a beam splitter. In the CASSI arm, a low-spatial resolution CA {modulates the spatial information, and afterwards a prism disperses} the coded information into a monochromatic sensor. In the MCFA arm, a low-spectral resolution CCA codifies the spatial-spectral information which is then integrated into a monochromatic sensor. }
    \label{sytems}
\end{figure}
\section{CSI observation model}\label{models}
First, we derive the discretization of the desired high-spatial-spectral resolution image in terms of the pitch size of each CSI architecture. In this work, we consider the CASSI architecture which provides high spectral resolution given by the spectral response of the prism with a spectral pitch size given by $\Delta_\lambda$ and a low spatial resolution corresponding to a pitch size $\Delta_S$. Also, we consider the MCFA  with a high spatial resolution $\Delta_s$ and a low spectral resolution due to spectral response of the filter $\Delta_\Lambda$ as is illustrated in Fig.\ref{sytems}. These factors are related as: 
 \begin{align}
     \Delta_S = d_s\Delta_s,\\
     \Delta_\Lambda = d_\lambda\Delta_\lambda,
 \end{align}
 where the parameters $d_s, d_\lambda$ are integer up-scaling factors. The discretization of the target high spatial spectral resolution image is formulated in terms of $\Delta_s$ and $\Delta_\lambda$ as follows
\begin{align}
    \mathbf{F}_{(m,n,\ell)} =  &\iiint f(x,y,\lambda)\nonumber\\&\times\text{rect}\Bigg(\frac{x}{\Delta_{s}} - m,\frac{y}{\Delta_{s}} - n, \frac{\lambda}{\Delta_{\lambda}} - \ell \Bigg)dxdyd\lambda,
    \label{voxels}
\end{align}
where $f(x,y,\lambda)$ is the continuous representation of the target scene, $m=1,\dots, M$, $n=1,\dots, N$ and $\ell=1,\dots, L$ and $M,N$ and $L$ are the spatial-spectral dimensions.

\subsection{CASSI Forward  Sensing Model} \label{CASSI}

In the CASSI architecture, the input light source is first focused by imaging lens to a CA, which codifies the spatial information of the image. The CA can be implemented in spatial light modulators (SLM) such as digital micromirror devices (DMD)\cite{DMD} which block/unblocks certain regions of the scene. Then, the spectral information of the coded field is dispersed through a prism. Finally, the coded and dispersed information impinges a focal plane array, as shown in one arm of the proposed system in Fig.\ref{sytems}. Since this sensing architecture is modeled as a low-spatial resolution system, a decimation operation is present in the mathematical model. Therefore, the  discrete model of the CASSI measurements $\mathbf{g}_c$ can be formulated as: 
\begin{align}
    \mathbf{g}_{c_{(i,j)}} =& \sum_{\ell = 1}^{L}\mathbf{H}_{c_{(i,j)}}\sum_{t=1+(i-1)d_s}^{id_s}\sum_{q=1+(j-1)d_s}^{jd_s} \mathbf{F}_{(i-t,j-q-\ell,\ell)},
\end{align}
where the summations in $t$ and $q$ represent the spatial resolution difference between the target high-spatial-spectral image $\mathbf{F}$ and the CASSI measurements,  $\mathbf{H}_c$ represents the coded aperture. Due to the dispersion response of the prism, a dispersed voxel impinges more than one pixel on the sensor, to model this effect, { in \cite{high-order} a high-order CASSI model is proposed}, which dictates that one sheared voxel affects up to three neighboring pixels on the FPA. The energy distribution due to this effect is modeled by the weights $\mathbf{Q}_{(i,j,\ell,u)}$ where $u=0,1,2$ stands for the region of the model. This model can be seen as a weighted average of the dispersed voxels in the sensor. Then, the discrete high-order CASSI model is given by 
\begin{align}
    \mathbf{g}_{c_{(i,j)}} = \sum_{\ell=1}^{L}\sum_{u=0}^{2}  \mathbf{H}_{c_{(i,j)}}\sum_{t}&\sum_{q} \mathbf{Q}_{(i,j,\ell,u)} \mathbf{F}_{(i-t,j-q-\ell-u,\ell)}. \label{cassi_discrete}
\end{align}
The discrete model in \eqref{cassi_discrete} can be expressed in a matrix-vector product in the following expression
\begin{equation}
    \mathbf{g_c} = \mathbf{\Phi}_c\mathbf{f} + \mathbf{n}_c,
\end{equation}
where $\mathbf{g}_c \in \mathbb{R}^{\tilde{M}(\tilde{N}+L-1)}$ are the compressed measurements, with $\tilde{M} = \frac{M}{d_s}$ and $\tilde{N} = \frac{N}{d_s}$, $\mathbf{\Phi_c}\in\mathbb{R}^{\tilde{M}(\tilde{N}+L-1)\times MNL}$ the CASSI sensing matrix, $\mathbf{f} \in \mathbb{R}^{MNL}$ is the vecotrization of the high spatial-spectral resolution image, and $\mathbf{n}_c \in \mathbb{R}^{\tilde{M}(\tilde{N}L-1)}$ is an additive noise. 
\begin{figure*}[t]
    \centering
    \includegraphics[width=\textwidth]{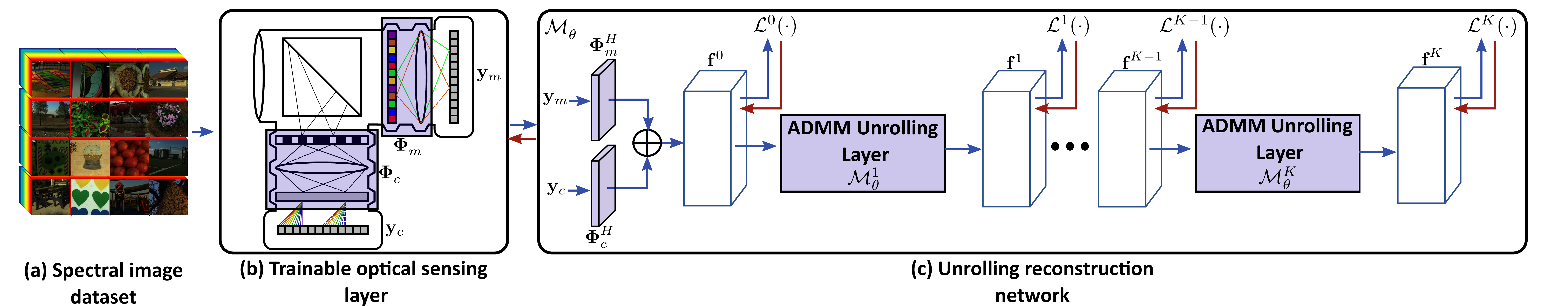}
    \caption{Proposed D$^\text{2}$UF framework, with a spectral image dataset (a), the parameters of the optics systems are jointly optimized. (b) The optical sensing systems are modeled as a layer where the CA and the CCA are the trainable parameters, and they are coupled with a reconstruction network (c). The reconstruction network is an ADMM-based unrolling network, where each layer of the network is interpreted as an ADMM step. Finally, the loss function is computed at the output of each unrolling stage to increase the performance and convergence of the early stages.}
    \label{fig:e2e}
\end{figure*}
\subsection{MCFA Forward  Sensing Model} \label{MCFA}

In the MCFA system, the input light source is focused into a CCA, which jointly codifies the spatial and the spectral information of the scene. {The coded field is then integrated} into an FPA detector as shown in one of the arm in Fig.~\ref{sytems}. This system is modeled as a low spectral resolution given by the resolution of the filter. Then, the discrete measurements on the FPA detector are given by 
\begin{equation}
    \mathbf{g}_{m_{(i,j)}} = \sum_{\ell = 1}^{\frac{L}{d_\lambda}}\mathbf{H}_{m_{(i,j,\ell)}}\sum_{p = 1+(\ell-1)d_\lambda}^{\ell d_\lambda} \mathbf{F}_{(i,j,\ell-p)},
\end{equation}
where $\mathbf{H}_m$ is the discretization of the CCA. The model can be also expressed a matrix-vector product as
\begin{equation}
     \mathbf{g}_m = \boldsymbol{\Phi}_m\mathbf{f} + \mathbf{n}_m,
\end{equation}
where $\mathbf{g}_m\in\mathbb{R}^{MN}$ is the compressive measurements,   $\mathbf{\Phi}_m \in \mathbb{R}^{MN\times MNL}$  is the sensing matrix and $\mathbf{n}_m\in\mathbb{R}^{MN}$ stands for the noise. 

A scheme of the CSIF systems based on the CASSI and MCFA optical architectures is shown in Fig. \ref{sytems} where a beam-splitter divides the incident light source to the  MCFA and the CASSI architectures.

\section{End-to-End formulation for compressive spectral image fusion}\label{e2e}
The E2E optimization requires an optical layer formulation of the sensing system considering fully differentiable modeling with respect to the optical parameters for an efficient update in the backpropagation scheme. Consequently, the following two sections present the forward and backward modeling of the CASSI and MCFA layers. This E2E methodology is depicted in Fig. \ref{fig:e2e}(b)
\subsection{Optical layer modeling of the sensing operators}
\subsubsection{CASSI}The sensing matrix of the CASSI architecture $\boldsymbol{\Phi}_c$ can be modeled as as: 
\begin{equation}
    \mathbf{\Phi}_c =\mathbf{P}\mathbf{T}_c \mathbf{D}_s,
\end{equation}
where $\mathbf{P} \in \mathbb{R}^{{\tilde{M}(\tilde{N}+L-1)\times \tilde{M}\tilde{N}L}}$  is a fixed matrix that models the dispersion effect considering  the high-order CASSI modeling, $\mathbf{T}_c\in \{0,1\}^{\tilde{M}\tilde{N}L\times \tilde{M}\tilde{N}L}$ is the diagonalized coded aperture matrix $\mathbf{H}_c$, and $\mathbf{D}_s \in \mathbb{R}^{\tilde{M}\tilde{N}L\times MNL}$ is a fixed spatial decimation operator that down-samples the measurements dimension by the factor $d_s$. The custom element in this layer is the CA. In our E2E approach, we propose to use a sigmoid activation function of the CA weights to encourage binary values on the resulting CA. Therefore, the $\mathbf{T}_c$ is formulated as: 
\begin{equation}
    \mathbf{T}_c =\operatorname{diag}\left(\sigma(\gamma_c\mathbf{W}_c)\right)
\end{equation}
where $\mathbf{W}_c\in\mathbb{R}^{\tilde{M}\times \tilde{M}}$ are the trainable parameters of the layer, $\sigma(\cdot)$ is the sigmoid, $\gamma_c$ is a hyperparameter that controls the smoothness of the sigmoid, the greater $\gamma$ the functions tends to a better binarization as shown in Fig. \ref{fig:sigmoid}. Then the forward model of the CASSI layer is given by

\begin{equation}
    \mathbf{g}_c =\mathbf{P}\operatorname{diag}\left(\sigma(\gamma_c\mathbf{W}_c) \right)\mathbf{D}_s\mathbf{f} + \boldsymbol{\omega}_c.
\end{equation}
Note that now we wish to learn the optimal values $\mathbf{W}_c$ since these weights generate the binary CA.

\subsubsection{MCFA} In this system the sensing process is described~as
\begin{equation}
    \mathbf{\Phi}_m = \mathbf{T}_m\mathbf{D}_\lambda,
\end{equation}
where $\mathbf{D}_\lambda \in \mathbb{R}^{MN\tilde{L}\times MNL}$ is a fixed spectral decimation operator which induces the up-scaling factor $d_\lambda$ in the spectral dimension, $\tilde{L} = \frac{L}{d_\lambda}$, $\mathbf{T}_m \in\{0,1\}^{MN\times MN\tilde{L}}$ is a trainable matrix that contains the diagonalization of the CCA entries $\mathbf{H}_m$.

Similarly, we used the same sigmoid function to generate the matrix $\mathbf{H}_m$ \textit{i.e.} $\mathbf{H}_m = \sigma\left(\gamma_m \mathbf{W}_m\right)$ where $\mathbf{W}_m\in\mathbb{R}^{MN\times MN\tilde{L}}$. Therefore, the forward MCFA operator is given by: 
\begin{equation}
     \mathbf{g}_m =  \operatorname{diag}\left(\sigma\left(\gamma_m \mathbf{W}_m\right)\right)\mathbf{D}_\lambda\mathbf{f} +\boldsymbol{\omega}_m.
\end{equation} 
\subsection{Derivative of the optical layers} 
\begin{figure}[!t]
    \centering
    \includegraphics[width=\linewidth]{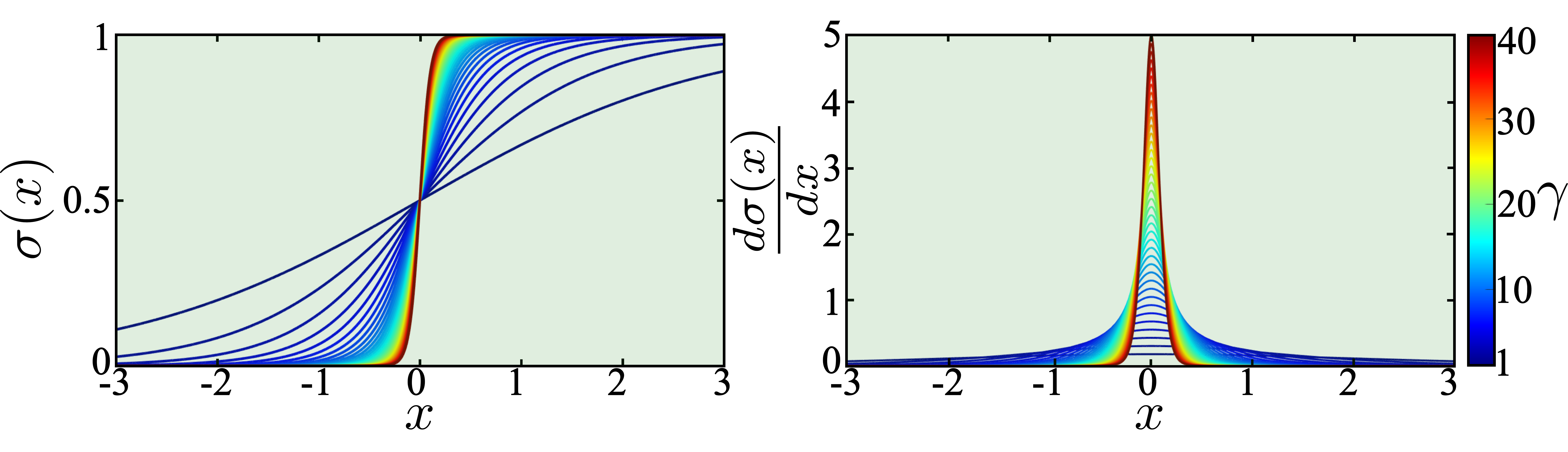}
    \caption{Effect of the $\gamma$ parameter in the sigmoid function employed for the activation function of the trainable optical parameters (left) and its derivative (right). Notice that the greater the value of $\lambda$ the sigmoid function tends to binarize better its inputs, but its derivative becomes sharper, {which produces that the parameters are barely updated via a gradient descent algorithm}.}
    \label{fig:sigmoid}
\end{figure}

An essential aspect of the sensing layer parameters training are its derivatives such that the trainable parameters can be optimized effectively. First, we can formulate the derivative of the CASSI with the chain rule: 
\begin{align}
    \frac{\partial \mathbf{g}_c}{\partial \mathbf{W}_c}=\frac{\partial \mathbf{g}_c}{\partial \mathbf{H}_c}\frac{\partial \mathbf{H}_c}{\partial \mathbf{W}_c} = \left(\mathbf{D}_s^T \otimes \mathbf{P}\right)\sigma'(\gamma_c\mathbf{W}_c),
\end{align}
here, a key aspect is the $\gamma_c$ parameter since the greater $\gamma_c$, the sharper the derivative becomes, and the weights $\mathbf{W}_c$ are barely updated through gradient descent based algorithm. Fig. \ref{fig:sigmoid} shows the sigmoid function for different $\gamma$ values where the greater this value becomes, the more binary the function is, but its derivative is sharper. 
A similar analysis is made for the MCFA forward model, the {derivative of which is described as}: 
\begin{equation}
    \frac{\partial \mathbf{g}_m}{\partial \mathbf{W}_m}=\frac{\partial \mathbf{g}_m}{\partial \mathbf{H}_m}\frac{\partial \mathbf{H}_m}{\partial \mathbf{W}_m} = \left(\mathbf{D}_s^T\right)\sigma'(\gamma_m\mathbf{W}_m),
\end{equation}

which is also controlled by the value of the $\gamma_m$ parameter.
\subsection{E2E Optimization}
Modeling the acquisition systems as optical layers, the E2E optimization  for CSIF with a training dataset of $T$ spectral images is formulated as
\begin{align}
    \{\hat{\boldsymbol{\theta}},\hat{\boldsymbol{\Phi}}_m,\hat{\boldsymbol{\Phi}}_h\} = \mathop{\arg\min}\limits_{\boldsymbol{\theta},\boldsymbol{\Phi}_m,\boldsymbol{\Phi}_h} &\frac{1}{T}\sum_{t=1}^{T}\mathcal{L}(\mathcal{M}_{\theta}\left(\boldsymbol{\Phi}_m\mathbf{f}_t,\boldsymbol{\Phi}_h\mathbf{f}_t\right),\mathbf{f}_t) + \nonumber\\& \mu_b (R(\mathbf{W}_m)+R(\mathbf{W}_h)),
    \label{eq_E2E}
    \end{align}
where $\mathcal{M}_{\boldsymbol{\theta}}$ is the reconstruction network with trainable parameters $\boldsymbol{\theta}$. The last term in the loss function is a regularization function to encourage binary values on the CA and CCA of the sensing layer. The main reason for employing these regularization functions is to relax the sigmoid function (\textit{i.e.} decrease $\gamma$) to obtain a smoother derivative while still obtaining binary values. The regularization function employed is the one proposed in \cite{bacca2021deep}:
\begin{equation}
    R(\mathbf{W}) = \sum_{i,j}(\sigma\left(\gamma\mathbf{W}_{(i,j)}\right))^2(\sigma\left(\gamma\mathbf{W}_{(i,j)}\right)-1)^2,
\end{equation}
where this function is minimized at $0$, and $1$. Therefore, we control the training of the CA and CCA by the sigmoid function and with the regularizer. For this purpose, a good choice of the regularization parameter $\rho_b$ and sigmoid parameters $\gamma_c$ and $\gamma_m$ allows an optimal training of the sensing parameters. 
\subsection{Proposed CSI Loss Function}
For CSI recovery, {the use of the mean squared error (MSE) as a loss function is widespread  \cite{dnu,unr3,lambda_net}}. However, these approaches do not exploit the data cube's spatial and spectral structural information. To address this issue, we propose a loss function  composed of two parts, spatial ($\mathcal{L}_s$) and spectral ($\mathcal{L}_\lambda$) losses, $ \mathcal{L} = \mu_s\mathcal{L}_{s} + \mu_\lambda\mathcal{L}_{\lambda}$, where $\mu_s$ and $\mu_\lambda$ are weighting hyperparameters. The spatial loss function $\mathcal{L}_s$ is inspired in the one proposed in \cite{loss} and it is defined, for the $t$-th training image, as:
\begin{equation}
\small
    \mathcal{L}_{s} = \frac{1}{T}\sum_{t=1}^{T}\frac{1}{MNL}\left(  1-\text{MS-SSIM}(\mathbf{f}_t,\hat{\mathbf{f}}_{t})+||\mathbf{f}_t-\hat{\mathbf{f}}_t||_1\right),
\end{equation}
the mixture of the $\ell_1$-norm allows color preservation and illuminance fidelity,  and the multiscale structural similarity (MS-SSIM) addresses the high-frequency contrast. The spectral loss minimizes the angle between the reconstructed and ground-truth spectral signatures, a valued parameter for SI applications. To derive the final expression of the spectral loss function, first define the angle $\beta$ of the estimated $i$-th spectral signature of the $t-$th ground truth image $\mathbf{f}_{t_{(i)}}$ and its estimated signature $\hat{\mathbf{f}}_{t_{(i)}}$ as 
\begin{equation}
    \beta = \cos^{-1}\left( \frac{\mathbf{f}_{t_{(i)}}^T\hat{\mathbf{f}}_{t_{(i)}}}{\Vert\mathbf{f}_{t_{(i)}}\Vert_2\Vert\hat{\mathbf{f}}_{t_{(i)}}\Vert_2} \right),
\end{equation}
the angle $\beta$ is minimized when the argument of the $\cos^{-1}$ tends to $1$ \textit{i.e.},
\begin{equation}
  \mathbf{f}_{t_{(i)}}^T\hat{\mathbf{f}}_{t_{(i)}}=\Vert\mathbf{f}_{t_{(i)}}\Vert_2\Vert\hat{\mathbf{f}}_{t_{(i)}}\Vert_2,
\end{equation}
We  formulate the spectral loss function as follows:
\begin{equation}
    \mathcal{L}_{\lambda} = \frac{1}{T}\sum_{t=1}^T\frac{1}{MN}\sum_{i}^{MN}\left( \mathbf{f}_{t_{(i)}}^T\hat{\mathbf{f}}_{t_{(i)}} - \Vert\mathbf{f}_{t_{(i)}}\Vert_2\cdot\Vert\hat{\mathbf{f}}_{t_{(i)}}\Vert_2\right)^2,\vspace{-0.1em}
\end{equation}
In this sense, our loss function promotes the direct enhancement of the visual perception of the reconstructed image jointly with the spectral fidelity of the reconstructed spectral image.
\section{Unrolled Reconstruction Network}\label{recon}
This paper proposes to adapt a traditional optimization approach to a data-driven scheme, where the DNN is interpreted as an iterative optimization algorithm. To give optimization-based interpretability to the reconstruction network, referred to as $\mathcal{M}_\theta(\cdot)$, first formulate the CSIF reconstruction process as the following optimization problem
\begin{equation}
    \label{opt1}
    \hat{\mathbf{f}} = \mathop{\arg\min}\limits_{\mathbf{f}}\frac{1}{2}||\boldsymbol{\Phi}_c\mathbf{f}-\mathbf{y}_c||_2^2 + \frac{1}{2}||\boldsymbol{\Phi}_m\mathbf{f}-\mathbf{y}_m||_2^2+\tau R(\mathbf{f}), 
\end{equation}
where $R(\mathbf{f})$ is the regularization function that promotes prior information of the spectral image such as sparsity \cite{CSI}, low-rank \cite{lr} or total variation \cite{CSIF1}. The problem in \eqref{opt1} can be solved using ADMM. First, introducing an auxiliary variable $\mathbf{h}\in \mathbb{R}^{MNL}$ we have:  
\begin{align}
    \{\hat{\mathbf{f}},\hat{\mathbf{h}}\} =  \mathop{\arg\min}\limits_{\mathbf{f},\mathbf{h}}&\frac{1}{2}||\boldsymbol{\Phi}_c\mathbf{f}-\mathbf{y}_c||_2^2 + \frac{1}{2}||\boldsymbol{\Phi}_m\mathbf{f}-\mathbf{y}_m||_2^2+\tau R(\mathbf{h}) \nonumber \\&\text{subject to } \mathbf{f=h},\label{admm1}
\end{align}
then ADMM aims to minimize the augmented Langrangian of the constrained optimization problem: 
\begin{align}
    \mathcal{L}(\mathbf{f,h,u}) = & \frac{1}{2}\Vert\boldsymbol{\Phi}_c\mathbf{f}-\mathbf{y}_c\Vert_2^2+\frac{1}{2}\Vert\boldsymbol{\Phi}_m\mathbf{f}-\mathbf{y}_m\Vert_2^2\nonumber\\&+\tau R(\mathbf{h}) + \frac{\rho}{2}\Vert\mathbf{f} - \mathbf{h} + \mathbf{u}\Vert_2^2,\label{lagrangian}
\end{align}
where $\mathbf{u}\in \mathbb{R}^{MNL}$ is the Lagrange multiplier. The solution of \eqref{lagrangian}  is found solving the following sub-problems
\begin{align}
    \mathbf{h}^{k+1} = \underset{\mathbf{h}}{\textrm{argmin }} \tau R(\mathbf{h}) + \frac{\rho}{2}\Vert\mathbf{f}^{k} - \mathbf{h} + \mathbf{u}^{k}\Vert_2^2 \label{h_problem}
    \end{align}
    \begin{align}
    \mathbf{f}^{k+1} = \underset{\mathbf{f}}{\textrm{argmin }}& \frac{1}{2}\Vert\boldsymbol{\Phi}_c\mathbf{f}-\mathbf{y}_c\Vert_2^2+\frac{1}{2}\Vert\boldsymbol{\Phi}_m\mathbf{f}-\mathbf{y}_m\Vert_2^2\nonumber\\& + \frac{\rho}{2}\Vert\mathbf{f} - \mathbf{h}^{k+1} + \mathbf{u}^{k}\Vert_2^2\label{f_problem}
    \end{align}
    \begin{align}
    \mathbf{u}^{k+1} = \mathbf{u}^{k} + \alpha^k(\mathbf{f}^{k+1}-\mathbf{h}^{k+1}).\label{u_problem}
\end{align}

\begin{figure}[t]
    \centering
    \includegraphics[width=\linewidth]{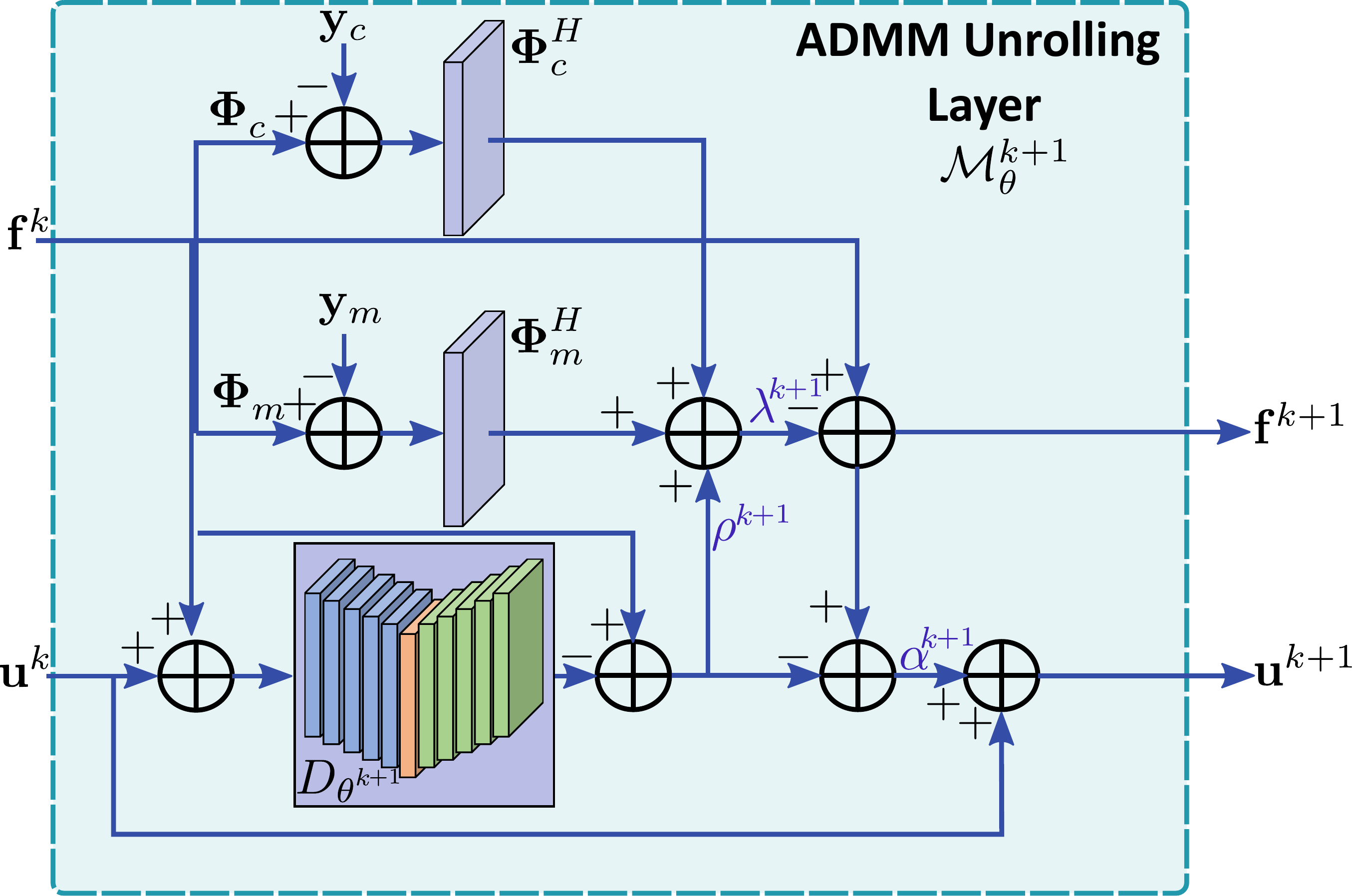}
    \caption{Operations of the $\mathcal{M}_\theta^k$ layer  where the trainable parameters are the optimization parameters $\lambda^k,\rho^k$ and $\alpha^k$ as well as deep prior network which performs a proximal mapping.}
    \label{fig:unrolling}
\end{figure}

The sub-problem \eqref{h_problem}  can be solved using a proximal operator of the image prior. For instance, if the prior is the sparsity, the proximal operator can be a hard-thresholding operator \cite{IHT}, or a soft-thresholding opertator \cite{FISTA}. Here, a deep prior is employed, {which only requires learning a proximal solver through a subnetwork in the unrolling network instead of learning the prior explicitly}. Therefore, denote $D_{\theta^k}(\cdot)$ the deep proximal operator where $\theta^k$ are parameters of the CNN in the iteration k. Consequently, the iterations of $\mathbf{h}$ are:
\begin{equation}
    \mathbf{h}^{k+1} = D_{\theta^{k+1}}(\mathbf{f}^{k}+\mathbf{u}^k) .
\end{equation}
Several networks have been proposed for this proximal step. For instance, \cite{unr1} proposed a deep prior network composed of a spatial network and a spectral network, \cite{dnu} designed a network based on local and non-local relations of the image, and \cite{zhang2021learning} exploits low-rank representation of the image. Then, the problem \eqref{f_problem} is solved using a gradient descent method in which each step is given by:
\begin{align}
\mathbf{f}^{k+1} = \mathbf{f}^{k} -& \lambda^{k+1}\left[\boldsymbol{\Phi}_c^T\left(\boldsymbol{\Phi_c}\mathbf{f}^{k}-\mathbf{y}_c\right)+\boldsymbol{\Phi}_m^T\left(\boldsymbol{\Phi_m}\mathbf{f}^{k}-\mathbf{y}_m\right)\right.\nonumber\\&\left.+\rho^{k+1}\left(\mathbf{f}^{k}-\mathbf{h}^{k+1}+\mathbf{u}^k\right)\right].
\end{align}
We can represent each ADMM step as a layer of the reconstruction network such that
\begin{equation}
    \{\mathbf{f}^{k+1},\mathbf{u}^{k+1}\} = \mathcal{M}_\theta^{k+1}(\mathbf{f}^k,\mathbf{u}^k),
\end{equation}
with trainable parameters $\lambda^{k+1}, \alpha^{k+1}$, $\rho^{k+1}$ and the parameters of the prior network $D_\theta^{k+1}(\cdot)$. Then $K$ iteration layers are unfolded in a DNN denoted as $\mathcal{M}_{\theta}(\cdot)$. Figure~\ref{fig:unrolling} shows the structure of the interpretable layer $\mathcal{M}_\theta^k(\cdot)$ in the unrolled network, where the proximal operator is learned during the training stage. In our approach,  $\mathbf{f}^0 = \frac{1}{2} \left(\boldsymbol{\Phi}_c^T\mathbf{y}_c+\boldsymbol{\Phi}_m^T\mathbf{y}_m\right)$ and $\mathbf{u}^0 = \mathbf{0}$ are used as initialization. 
\subsection{Multiple Loss Strategy}

{The unrolling networks are typically very deep, which causes} the vanishing of the gradient in the first layers, yielding a poor estimation of the spectral images at these stages\cite{unr1}. For this reason, we propose to compute the loss function at the end of each unrolling stage, {such that the} loss is backpropagated more uniformly in the entire network, improving the reconstruction quality of the first stages. Therefore, the loss function at the end of each stage is described as
\begin{equation}
    \mathcal{L}^k = \mu_s\mathcal{L}_s(\hat{\mathbf{f}}_s^k,\mathbf{f}_s) +\mu_s\mathcal{L}_\lambda(\hat{\mathbf{f}}_s^k,\mathbf{f}_s),
\end{equation}
where $\hat{\mathbf{f}}_s^k=\mathcal{M}_{\theta}\left(\boldsymbol{\Phi}_m\mathbf{f}_s,\boldsymbol{\Phi}_h\mathbf{f}_s\right),\mathbf{f}_s)$ is the spectral image estimated at the $k-$th stage. The total loss function is denoted~as 
\begin{equation}
    \mathcal{L} = \sum_{k=0}^{K}\mathcal{L}^k(\hat{\mathbf{f}}_s^k,\mathbf{f}_s).\label{multi_loss}
\end{equation}
Additionally, the loss function is computed in the initialization stage $\mathcal{M}_\theta^0 =\frac{1}{2} \left(\boldsymbol{\Phi}_c^T\mathbf{y}_c+\boldsymbol{\Phi}_m^T\mathbf{y}_m\right)$, allowing for a better update of sensing layers parameter to obtain a better initial estimation with only the transpose operation of the sensing layers. This allows data-driven regularization on both optical trainable parameters which directly improves the conditioning of the sensing matrix \cite{bacca2021deep}.
\begin{figure}[!b]
    \centering
    \includegraphics[width=\linewidth]{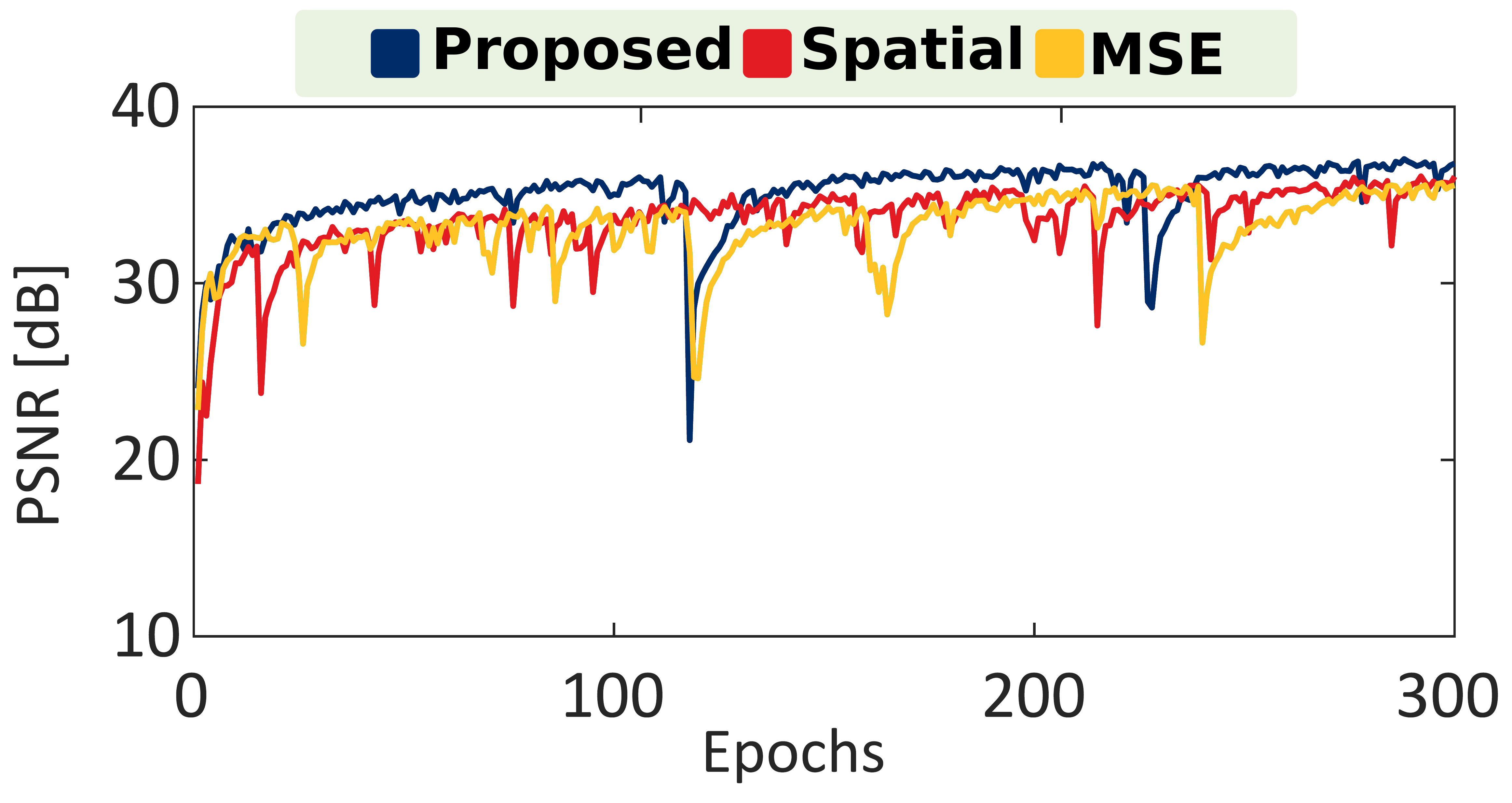}
    \caption{Validation performance during training for the ICVL dataset with different loss functions for 10 unrolling stages.}
    \label{fig:loss}
\end{figure}

\section{Simulation results}\label{sims}

To validate the proposed method, numerical simulations were conducted using two well-known spectral image datasets; the ICVL dataset~\cite{icvl} which has 180 spectral images, split into 140 for training, 20
for validation, and 20 for testing, and the ARAD dataset~\cite{arad_challenge} with 460 spectral images divided into 400 for training, 30 for
validation, and 30 for testing. The results shown for the proposed method in each section were obtained using the ADAM optimizer~\cite{adam} for 300 epochs, reducing the learning rate every 50 epochs to a half. The peak signal-to-noise ratio (PSNR)~\cite{5596999}, structural similarity index (SSIM)~\cite{ssim}, and the SAM \cite{sam1} metrics were used to evaluate the recovery performance. All simulations were performed in a GPU NVIDIA RTX 3090.

\subsection{Ablation study of the loss function}
We compare the performance of the proposed loss function with the mean square error (MSE) and using only the spatial loss function $\mathcal{L}_s$. For this simulation, 10 unrolling stages were employed using a CA and CCA fixed randomly with a transmittance of 0.5. Furthermore, the ICVL dataset and $d_s=d_\lambda=2$ were used as dataset and decimator factors, respectively. For the proposed loss function, the weighting coefficients were set as $\rho_s=\rho_\lambda=1$.
\begin{figure}[!t]
    \centering
    \includegraphics[width=0.95\linewidth]{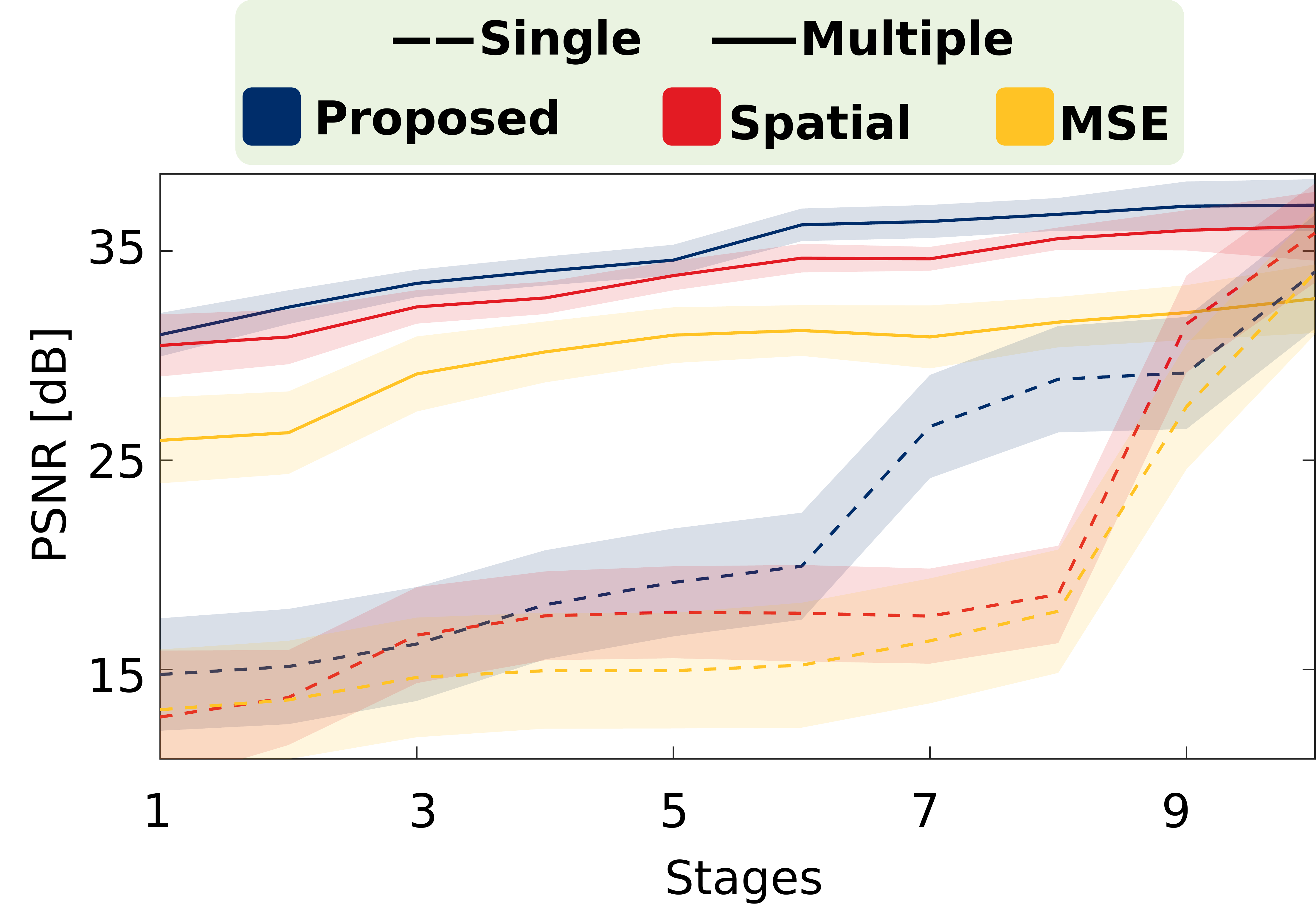}
    \caption{Validation performance of the network computing the loss function at the end of each stage and only at the end of the network, for the proposed, spatial, and MSE loss function.}
    \label{fig:multi_loss}
\end{figure}

First, the performance of the network was analyzed for the mentioned loss function during training.  Figure~\ref{fig:loss} shows the performance at the end of the unrolling network for the validation dataset. The obtained results show that the proposed loss function, which combines both spatial and spectral information, gives a better performance than the MSE and the spatial loss function in all the training processes. 
\begin{table}[!b]
\resizebox{0.49\textwidth}{!}{\begin{tabular}{cc|c|c|ccc}\toprule[2pt]
\multicolumn{2}{c|}{\textbf{Optical layer optimization}}                   & \multirow{2}{*}{\textbf{\begin{tabular}[c]{@{}c@{}}Sigmoid \\ activation\end{tabular}}} & \multirow{2}{*}{\textbf{\begin{tabular}[c]{@{}c@{}}Dynamic \\ parameters\end{tabular}}}    & \multicolumn{3}{c}{\textbf{Performance}}                                                           \\ \cline{1-2} \cline{5-7} 
\textbf{MCFA}                                   & \textbf{CASSI}            &                                                                                         & & \textbf{PSNR {[}dB{]}$\uparrow$} & \textbf{SSIM $\uparrow$}    & \textbf{SAM{[}rad{]}$\downarrow$}  \\ \toprule[2pt]
X                                               & X                         & \checkmark          & X                               & 36.12                            & 0.9609                      & 0.1423                            \\ \hline
X                                               & \checkmark & \checkmark  &X                                       & 37.8                             & 0.9651                      & 0.1125                            \\ \hline
\checkmark                       & X                         & \checkmark      &X                                      & 37.2                             & 0.9590                      & 0.1423                            \\ \hline
\checkmark                       & \checkmark & X                                                       &X             & 39.45                            &0.9792                  & 0.1224                           \\ \hline
{\checkmark} & \checkmark & \checkmark  &X                                          &\cellcolor{blue!12} 40.17  &  \cellcolor{blue!12}{0.9812} & \cellcolor{blue!12}0.0952    
\\ \hline
{\checkmark} & \checkmark & \checkmark&  \checkmark                                          & \cellcolor{green!12}{41.86}    & \cellcolor{green!12} {0.9852} & \cellcolor{green!12}{0.0794}                           
\\ \toprule[2pt]
\end{tabular}}
\caption{Ablation study of the E2E optimization of the CASSI and MCFA optical layers and the effect of using the sigmoid activation function to binarize the trained CA and CCA. Also the importance of dynamically update of the binarization parameters $\mu_b$ and $\gamma$. The highlighted green values are the best performance and the blue ones are the second best. }
\label{e2e_table}
\end{table}
\begin{table*}[t!]
\resizebox{1\textwidth}{!}{\begin{tabular}{c|c|ccc|ccc|ccc|ccc|ccc}\toprule[2pt]
\multirow{2}{*}{Dataset} & \multirow{2}{*}{Method} & \multicolumn{3}{c|}{SNR = 20 {[}dB{]}} & \multicolumn{3}{c|}{SNR = 25 {[}dB{]}} & \multicolumn{3}{c|}{SNR = 30 {[}dB{]}} & \multicolumn{3}{c|}{SNR = 35 {[}dB{]}} & \multicolumn{3}{c}{None} \\
                         &                         & PSNR        & SSIM        & SAM       & PSNR       & SSIM        & SAM        & PSNR        & SSIM       & SAM        & PSNR       & SSIM        & SAM        & PSNR   & SSIM   & SAM    \\\toprule[2pt]
\multirow{5}{*}{ICVL}    & SIFCM                   &       23.51      &     0.504        &     0.407      &    27.02        &      0.682      &    0.2885        &     30.12        &      0.842      & 0.185           &     32.57       &       0.886      &  0.1183          &    34.21    &  0.928      &  0.077   \\
                         & SIFCM-D                 &       22.99      & 0.414            &     0.3700      &   27.92         &     0.624       &   0.223         &  30.05           &         0.757   &    0.1637        &  33.05          &    0.867         &    0.109        &    35.16    &    0.94     &     0.082   \\
                         & LADMM-Net               &      27.95       &     0.911        &      0.139     &    \cellcolor{blue!12}{29.83}       &  \cellcolor{blue!12}{0.931}           &     \cellcolor{blue!12}{0.128}       &    33.90         &       0.944     &     0.114       &     34.58       &    0.967         &     0.097       &   36.64    & 0.968      &   0.080     \\
                         & LADMM-Net-D             &     \cellcolor{blue!12}{ 28.53 }      &    \cellcolor{blue!12}{   0.918 }     & \cellcolor{blue!12}{ 0.132 }        &     29.82       &     0.929        &        0.129    &      \cellcolor{blue!12}{34.21}      &       \cellcolor{blue!12}{0.956}    &    \cellcolor{blue!12}{ 0.105}       &   \cellcolor{blue!12}{35.48}         &    \cellcolor{blue!12}{ 0.972}        & \cellcolor{blue!12}{0.091 }           &  \cellcolor{blue!12}{38.92}      &    \cellcolor{blue!12}{0.981}   &  \cellcolor{blue!12}{0.071}     \\
                         & \duf                & \cellcolor{green!12}33.91       & \cellcolor{green!12}0.8106      & \cellcolor{green!12}0.212     & \cellcolor{green!12}39.24      & \cellcolor{green!12}0.9547      & \cellcolor{green!12}0.1078     & \cellcolor{green!12}41.11       & \cellcolor{green!12}0.955      & \cellcolor{green!12}0.077      & \cellcolor{green!12}41.85      & \cellcolor{green!12}0.984       & \cellcolor{green!12}0.069      & \cellcolor{green!12}42.28  & \cellcolor{green!12}0.983  & \cellcolor{green!12}0.067  \\\toprule[2pt]
\multirow{5}{*}{ARAD}    & SIFCM                   &      21.74       &         0.355    &     0.370      & 25.66           &         0.557    &     0.252       &      28.83       &  0.7316          &    0.1746        &     31.19       & 0.844            &  0.124          &    34.45    &    0.926    &  0.112      \\
                         & SIFCM-D                 &        22.31    &    0.329         &    0.363       &          26.45  &    0.535        &    0.242        &    \cellcolor{blue!12} 30.13        &      0.727      &    0.156        &     \cellcolor{blue!12}33.26       &0.862            &     \cellcolor{blue!12}0.101      &  \cellcolor{blue!12}35.75      &    \cellcolor{blue!12}0.943    &     \cellcolor{blue!12}0.058   \\
                         & LADMM-Net               &       \cellcolor{blue!12}24.41      &     \cellcolor{blue!12} 0.811       &     \cellcolor{blue!12} 0.219     &     \cellcolor{blue!12}27.06       &   \cellcolor{blue!12}0.821          &    \cellcolor{blue!12}0.149        &      28.74       &       0.882     &     0.124       &       30.42     &     0.892        &  0.115          &    31.54    &  0.901      &    0.101    \\
                         & LADMM-Net-D              &       22.41      &      0.801       &      0.223     &     25.26       &   0.812          &    0.182        &      29.11       &       \cellcolor{blue!12}0.891     &     \cellcolor{blue!12}0.106       &       31.42     &     \cellcolor{blue!12}0.912        &  0.125          &    32.21    &  0.922      &    0.098        \\
                         & \duf                 & \cellcolor{green!12}37.28      & \cellcolor{green!12}0.936       & \cellcolor{green!12}0.074     &\cellcolor{green!12} 40.26      & \cellcolor{green!12}0.969      & \cellcolor{green!12}0.051     & \cellcolor{green!12}41.97       & \cellcolor{green!12}0.980      & \cellcolor{green!12}0.434      & \cellcolor{green!12}43.05      & \cellcolor{green!12}0.984      & \cellcolor{green!12}0.04       &\cellcolor{green!12} 43.79  & \cellcolor{green!12}0.987 & \cellcolor{green!12}0.0383\\\toprule[2pt]
\end{tabular}}
\caption{Comparison of the recovery spectral image for the ICVL and the ARAD datasets with compressed measurements corrupted by different levels of AWGN. The highlighted green values are the best performance and the blue ones are the second best.}\vspace{-0.5cm}
\label{noise_table}
\end{table*}

Then, we analyze the effect of evaluating the loss function at the end of each unrolling stage ( referred to as multiple loss) compared with evaluating the loss function at the end of the network, as in state-of-the-art unrolling networks (single loss). For this, we fixed the number of unrolling stages as 10, and we compared the validation performance in every stage for the models trained with respect to those with multiple and single loss functions. The results are summarized in Fig \ref{fig:multi_loss}. There, it can be seen that the multiple loss strategy allows the early stages to have almost twice the performance of the single loss models, thus requiring fewer stages to reach the optimal performance. Nevertheless, the performance at the end of the network is similar to both modalities, {with  the multiple loss methodology performing slightly better}. The number of states in the unrolled network is a costly hyperparameter since it is a trade-off between quality and time processing. The state-of-the-art methods set this parameter training the network several times for different stages using the single loss at the end of the state. Therefore,  multiple-loss training gives a practical guide on how many iterations are required for the fusion problem with a single training trial. For the rest of the paper, the proposed method was trained using ten stages and the multiple loss functions scheme.




\subsection{Ablation study of the end-to-end network}
This section evaluated the proposal to learn jointly the CAs and CCAs and the unrolling reconstruction network weights. To this end, three aspects were studied:

 \textbf{Optical layer optimization:} Here, the effect of the E2E methodology to train the optical parameters for each system in the fusion problem was evaluated. The results for this experiment are shown in the first rows of Table \ref{e2e_table}, where the best reconstruction performance is given when the CASSI and MCFA optical layers are joint optimized. Additionally, the training of only the sensing layer of the CASSI system performs better than the training only on the MCFA architecture, implying that this system has a greater effect in the reconstruction process.

\textbf{Sigmoid activation:} This experiment evaluated the binarization methodology using the sigmoid activation over the weights and the binary regularization, compared with the methodology present in \cite{bacca2021deep}, which only uses binary regularization. The parameters of the sigmoid and the regularization parameters were set $\gamma_c=\gamma_m=35$ and $\mu_b= 0.005$ respectively. The fourth and fifth rows of Table \ref{e2e_table} show these simulations in which both optical layers are optimized, highlighting that the proposed sigmoid activation and regularization leads to better reconstruction performance. 
     \begin{figure}[!b]
    \centering
    \includegraphics[width=\linewidth]{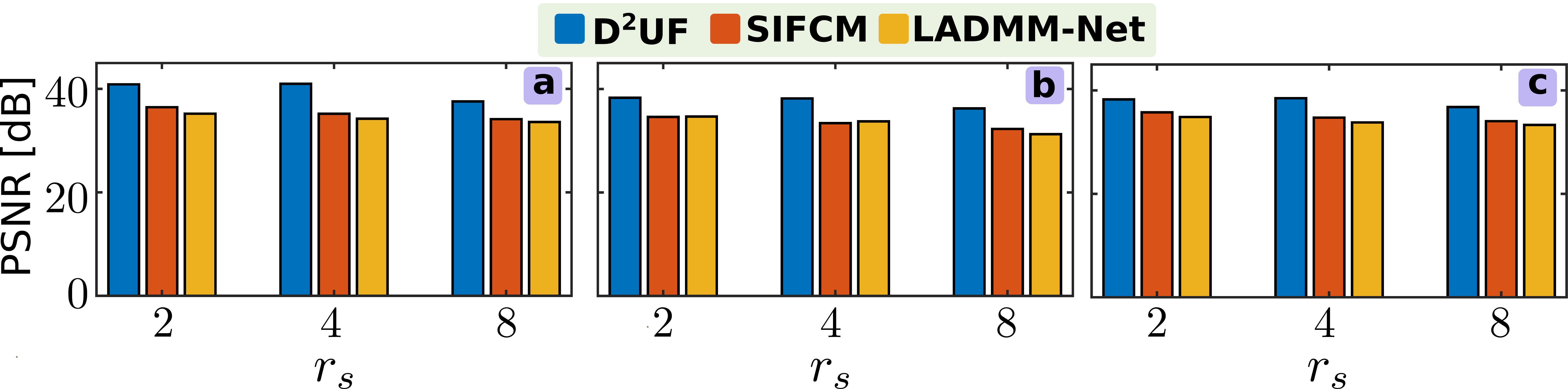}
    \caption{Performance of LADMM-Net, SIFCM and proposed method varying the spatial compression $r_s = {2,4,8}$ for $r_\lambda=2$ (a), $r_\lambda =4$ (b) and $r_\lambda=8$(c) for the ICVL dataset. }
    \label{fig:decimation}\vspace{-0.8cm}
\end{figure}

\textbf{Dynamic parameters:} As suggested in \cite{bacca2021deep}, the regularization value needs to be updated dynamically during training. Specifically, this value starts with a low value so that, in the early epochs, it guides the training and then increases the parameter to obtain binary values on the CA and CCA. We proposed to update the smoothing parameters $\gamma_c,\gamma_m$ and the regularization parameter $\rho_b$ following the function for $r=1,\dots,R$ epochs $\gamma^r = \gamma^{R}\tanh(\kappa_\gamma\lfloor\frac{r}{\beta_\gamma}\rfloor)$  , $\mu^r = \mu^{R}\tanh(\kappa_\mu\lfloor\frac{r}{\beta_\mu}\rfloor) $ where $\gamma^{R}, \mu^{R}$ are the maximum value of the parameters, $\kappa_\gamma$ and $\kappa_{\mu}$ are a delay factors which controls how fast the function achieves its maximum value and $\beta_\gamma, \beta_\mu$ defines the frequency of the update. The $tanh$ function and the suitable delay parameter allow better control of the increasing dynamic parameter~\cite{bacca2021deep}. Here, it was set the parameters $\mu^R = 35$ , $\beta_\gamma=\beta_\mu= 5$, $\kappa_\gamma=\kappa_\mu=0.01$. The results of this training methodology are shown in the last row of Table \ref{e2e_table} exhibiting a significant improvement of up to 1 dB in PSNR metric compared with a fixed parameter, i.e., without updating them. Therefore, for the next section, the best configuration that includes the E2E design of both systems using the dynamic parameters and sigmoid activation was used, and it is referred to as \duf.

\subsection{State-of-the-art Comparison}
In this section, the performance of \duf~was compared with state-of-the-art CSIF methods. A convex-optimization-based spectral image fusion from compressive measurements (SIFCM)~\cite{CSIF1}, that  is a method based on an ADMM formulation with sparsity and total variation priors, and a deep-learning approach (LADMM-Net)~ \cite{CSIF6} where an ADMM-inspired unrolled reconstruction network. Both methods were modified for the sensing systems CASSI and MCFA. The hyperparameter for both methods were chosen following the suggestion of the own authors. Furthermore, it is important to note that, in these methods, the sensing matrices are not optimized. Therefore, we also compared the SIFCM and LADMM-Net methods using the optimized sensing matrices obtained with our E2E method, denoted as SIFCM-D and LADMMM-Net-D.

\textbf{Spatial and spectral decimation factors:} The performance of the \duf~network was compared  with LADMM-Net and SIFCM methods, varying the decimation factors $r_s$ and $r_\lambda$ with values $2,4,8$. For that, we used the ICVL dataset. Figure~\ref{fig:decimation} shows the results of this experiment showing an improvement of the proposed method compared with the other methods for each decimation factor, obtaining an improvement of up~to~4~dB.
\begin{figure*}[!t]
    \centering
    \includegraphics[width=\textwidth]{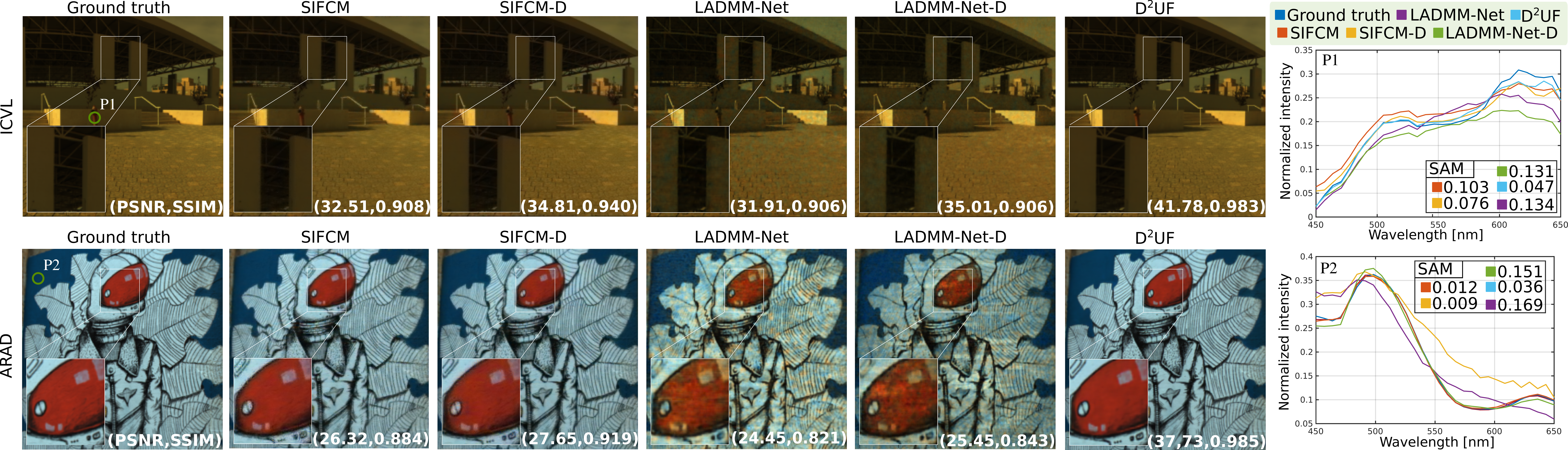}
    \caption{RGB representation of the reconstructed spectral images and the recovery of two representative spectral signature for a test image of the ICVL and ARAD dataset.}\vspace{-0.5em}
    \label{fig:visual}
\end{figure*}

 \begin{figure}[!b]
    \centering
    \includegraphics[width=0.49\textwidth]{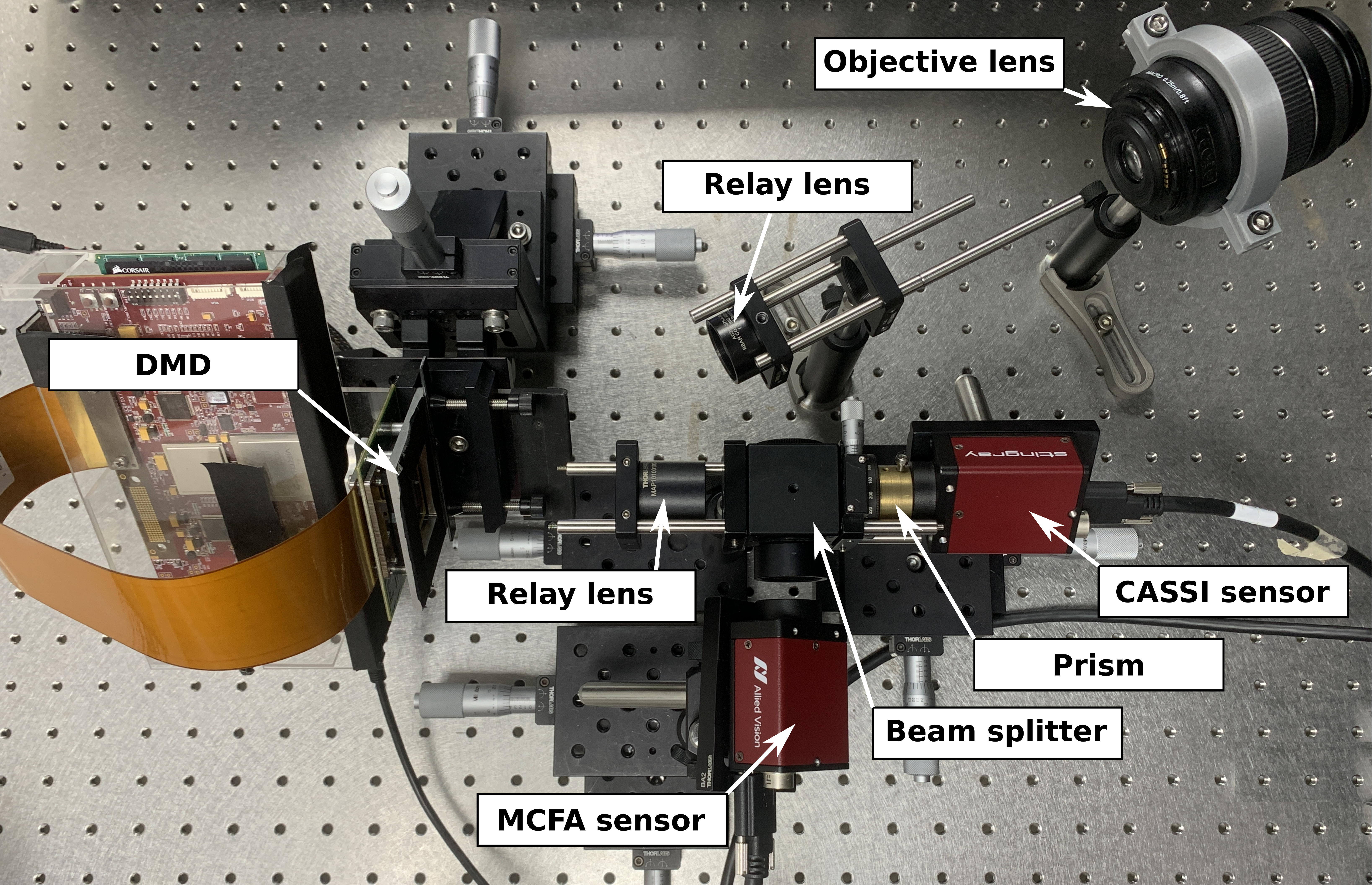}
    \caption{Experimental optical setup for real data validation.}
    \label{fig:exp_set_up}
\end{figure}

\textbf{Performance with noise:} In this experiment, we want to compare the performance of the proposed method with its counterparts in the presence of noise in compressed measurements. To this end, additive Gaussian noise was applied to vary the signal-to-noise-ratio (SNR) with $20,25,30,35$ [dB] and without noise. Here, {the ARAD and the ICVL dataset were used}. The results are shown in Table \ref{noise_table} {where it can be observed} that the proposed method outperforms the LADMM-Net and SCIFM algorithms. Remarkably, employing the designed CA and CCA in the LADMM-Net-D and SIFCM-D methods outperform the fusion quality regarding the random configuration, thus improving the conditionality of the fusion problem.

Finally, we present the visual results of the reconstructed spectral images. To this end,  an RGB representation of a reconstructed testing spectral image of the ICVL and the ARAD dataset employing the spec2rgb repository\footnote{\url{https://github.com/hdspgroup/spec2rgb.git}}is shown in Fig~\ref{fig:visual}. Remarkably, the proposed method preserves more highly detailed image features than its counterparts. Also, the reconstruction of two representative spectral signatures is shown {in which depicts the fact that the proposed method recovers high-fidelity spectral features}.

\section{Experimental results}\label{exp}

\begin{figure}[!t]
    \centering
    \includegraphics[width=\linewidth]{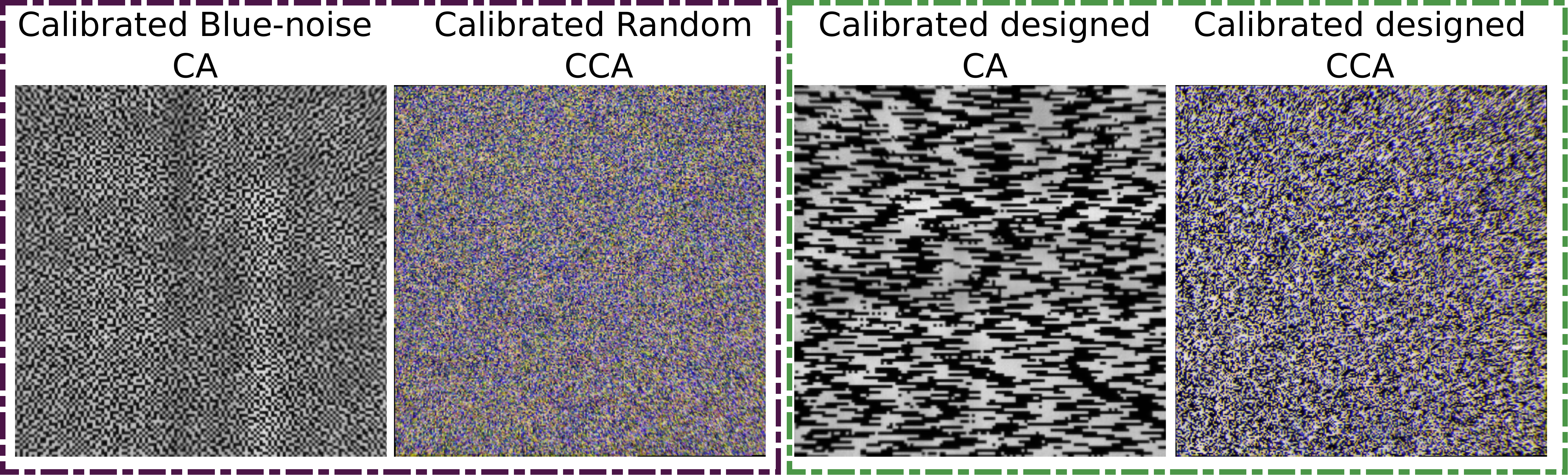}
    \caption{Calibrated sensing matrices for the blue noise CA, random CCA and the designed sensing matrices with the proposed method.}\vspace{-0.5em}
    \label{fig:calibrated}
\end{figure}

\begin{figure*}[!t]
    \centering
    \includegraphics[width=0.99\textwidth]{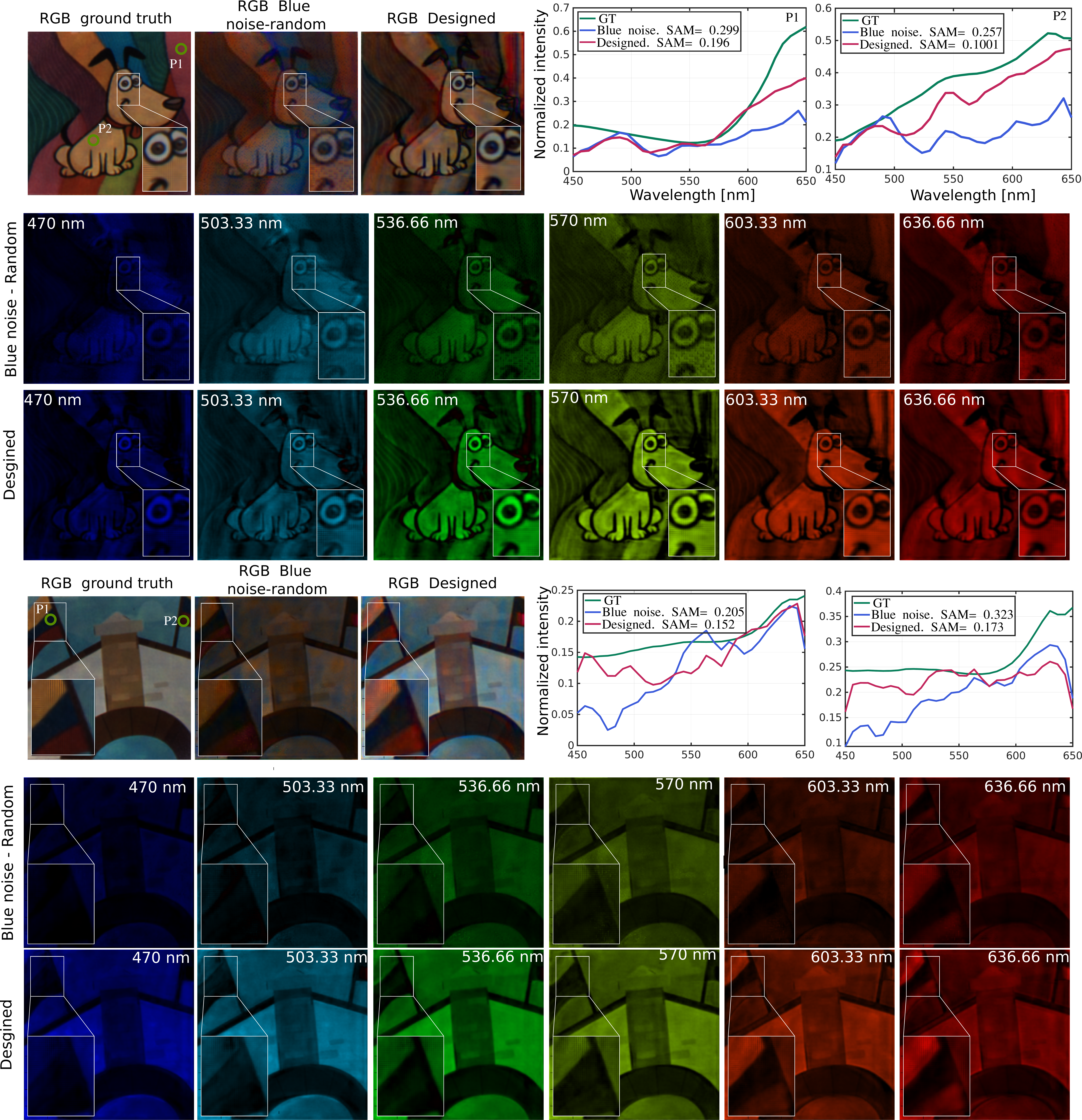}
    \caption{Reconstructed real data for two scenes in which it is compared the reconstruction of the proposed designed fusion method with a blue-noise design methodology. It is displayed 7  spectral bands for each science and the spectral reconstruction of two spectral signatures. Our method present better visual and spectral reconstruction that the comparison method.}
    \label{fig:real_data_recon}
\end{figure*}

This section evaluates the proposed method with experimental data obtained with an optical testbed implementation shown in Fig.~\ref{fig:exp_set_up}. The optical setup uses a Canon Macro 0.25m/0.8ft as an objective lens, two Achromatic Lens LSB08 Series Thorlabs of 100mm as relay lenses, and a non-polarizing Beam splitter CCM1-BS013 Thorlabs, an AMICI prism, a Texas Instruments DLP4130 DMD and two F-145 Stingray grayscale sensor. Here, the spatial resolution of the image was set to $512\times512\times31$ from the spectral range of 450 nm to 650 nm. The DMD was used for both the CASSI CA and the MCFA CCA. The CA resolution was set to $256\times 256$ and the CCA of $512\times512\times15$. For comparison purposes, compressive measurements were acquired using blue noise \cite{design1} design for the CA and a random CCA which are non-data dependent coding. Figure \ref{fig:calibrated} shows the calibrated CA and CCA obtained with the E2E design and blue-noise and random distribution. Our designed CA in CASSI exploits the shared patterns due to the prism dispersion and, in CCA, reduces the transmittance (more zero values), which helps to decrease information redundancy. Figure~\ref{fig:real_data_recon} shows an RGB mapping and $6/31$ spectral bands of two reconstructed scenes acquired with the experimental optical setup.
Furthermore, the mean of the spectral signature in a region of 10$\times$10 pixels is illustrated in Figure~\ref{fig:real_data_recon}. There it can be concluded that the proposed design's reconstruction is better than the blue noise and random sensing matrices even in real setups.
\section{Conclusion}
We present \duf,  a synergistic coded aperture optimization and recovery method for compressive spectral image fusion. The coded apertures design is done by modeling the fused systems as optical layers where the optical weights use the sigmoid activation function and binary regularization to constrain their values to implementable values. The optical layers are coupled with a proposed ADMM-based unrolling network that obtains the fused image. The coupled network is trained E2E via a multiple-loss strategy that efficiently determines the number of stages at the inference step since this methodology significantly improves the early and mid-stages of the network. The multiple-loss is based on the proposed loss function, promoting high spectral fidelity and a visual enhancement in the recovered image. Simulation results employing the proposed design with state-of-the-art fusion methods improve their performances compared with random distributions. Comparison with state-of-the-art compressive spectral fusion work shows that our methodology significantly outperforms these methods. Moreover, reconstructions with experimental data show that the design-coded apertures perform better than the blue-noise and random patterns.

\bibliographystyle{Transactions-Bibliography/IEEEtran.bst}
\bibliography{Biblio}
\begin{IEEEbiographynophoto}{Roman Jacome}
received the B.S degree in electrical engineering in 2021 from Universidad Industrial de Santander, Bucaramanga, Colombia. He is currently pursuing his master's degree in applied mathematics at Universidad Industrial de Santander. His interests include computational imaging, signal processing, hyperspectral imaging, and compressed sensing. 

\end{IEEEbiographynophoto}

\begin{IEEEbiographynophoto}{Jorge Bacca}
received the B.S. degree in computer science and the Ph.D. degree in Computer Science from the Universidad Industrial de Santander (UIS), Bucaramanga, Colombia, in 2017 and 2021, respectively. He is currently a Postdoctoral Researcher at UIS. He is a consulting associate editor for the IEEE Open Journal of Signal Processing. He is an author of more than 10 journal papers and 20 proceedings conferences related to optical design and high-level tasks directly to raw optical measurements. His current research interests include inverse problems, deep learning methods, optical imagining, and hyperspectral imaging.
\end{IEEEbiographynophoto}


\begin{IEEEbiographynophoto}{Henry Arguello}
received the Master’s degree in electrical engineering from the Universidad Industrial de Santander, Colombia, in 2003, and the Ph.D. degree from the Electrical and Computer Engineering Department, University of Delaware in 2013. In 2020 he was a visitor professor with the Computational Imaging group in Stanford. He is currently a Titular Professor with the Systems Engineering Department, Universidad Industrial de Santander, Bucaramanga, Colombia. He is a senior member of IEEE and Optical Society. He is associate Editor for the IEEE Transactions on Computational Imaging and the president of the signal processing chapter, Colombia. He was the Co-Chair and Technical Co-Chair of several international conferences and workshops. He is an author of more than 100 journal papers based on computational imaging
techniques, high dimensional signal coding and processing and optical design.\end{IEEEbiographynophoto}


\vfill


\end{document}